\documentclass[10pt,pre,bibnotes,nofootinbib,byrevtex,superscriptaddress,showpacs]{revtex4} 
\usepackage{amssymb}
\usepackage{graphicx} 

\newcommand{\be}{\begin{equation}} 
\newcommand{\ee}{\end{equation}} 
\newcommand{\sign}{\sigma} 
\newcommand{\erf}{\mbox{erf}} 
\newcommand{\erfc}{\mbox{erfc}} 
\newcommand{\corr}{\mathrm{corr}}

\begin{document} 
   
\title{Brownian motion with dry friction: Fokker-Planck approach} 
 
\author{Hugo Touchette} 
\email{h.touchette@qmul.ac.uk} 
\affiliation{\mbox{School of Mathematical Sciences, Queen Mary University of London, London E1 4NS, UK}} 
 
\author{Erik Van der Straeten} 
\email{e.straeten@qmul.ac.uk} 
\affiliation{\mbox{School of Mathematical Sciences, Queen Mary University of London, London E1 4NS, UK}} 
 
\author{Wolfram Just} 
\email{w.just@qmul.ac.uk} 
\affiliation{\mbox{School of Mathematical Sciences, Queen Mary University of London, London E1 4NS, UK}} 

\begin{abstract} 
We solve a Langevin equation, first studied by de~Gennes, 
in which there is a solid-solid or dry friction 
force acting on a Brownian particle in addition to the viscous 
friction usually considered in the
study of Brownian motion. We obtain both the time-dependent 
propagator of this equation 
and the velocity correlation function by solving the associated time-dependent 
Fokker-Planck equation. Exact results 
are found for the case where only dry friction acts on the particle. 
For the case where both dry 
and viscous friction forces are present, series representations of 
the propagator and correlation function 
are obtained in terms of parabolic cylinder functions. Similar series
representations
are also obtained for the case where an external constant 
force is added to the Langevin equation. 
\end{abstract} 


\maketitle
  
\section{Introduction} 
 
In one of his last papers, P.-G. de~Gennes~\cite{gennes2005} proposed to 
study Brownian motion under the 
influence of a solid-solid or dry friction force\footnote{Dry friction 
is also called Coulomb 
friction \cite{bowden1950}.} in addition to the viscous friction 
force commonly studied since the works of Einstein 
and Langevin \cite{gardiner1985}. The main property of dry friction, as is known from 
common experience, is that a certain 
threshold force $\Delta_F>0$ has to be applied to a solid object 
resting on a solid surface in order to 
move the object \cite{bowden1950}. If the applied force $F$ is 
smaller than $\Delta_F$, then the object stays at rest 
(stick state), whereas if $F>\Delta_F$, then the object starts moving (slip state). 
 
The simplest way to model this behaviour mathematically is to consider 
a velocity-dependent force having 
the form $-\Delta_F\sign(v)$, where $\sign(v)$ stands for the 
sign of $v$ with the convention $\sign(0)=0$. By incorporating this 
force into Newton's equation, 
\be 
m\dot v=-\Delta_F\sign(v)+F, 
\label{eqn1} 
\ee 
it is indeed easy to see that $v=0$ is the unique attracting state 
when $|F|< \Delta_F$, and that $v=0$ ceases to be attracting 
when $|F|\geq\Delta_F$. An object of mass $m$ initially at 
rest, $v(0)=0$, will therefore remain at rest 
unless an external force $F$ greater in magnitude than the threshold or 
contact force $\Delta_F$ is applied. 
 
This simple description of dry friction is commonly used in engineering 
and physics to study macroscopic systems
in which solid-solid friction plays a significant role 
\cite{olsson1998,persson1998,piedboeuf2000,berger2002}. It is used, 
for example, to study stick-slip 
motion \cite{elmer1997,leine1998,feng2003,stein2008}, vibrated 
granular media \cite{murayama1998,kawarada2004}, as well as the ratchet 
motion of solid objects moving on a surface which is vibrated 
horizontally by a 
time-dependent force \cite{buguin2006} (see also \cite{fleishman2007}). 
In this context, de~Gennes \cite{gennes2005} proposed to study the case where the 
surface vibrates randomly by considering 
$F$ in (\ref{eqn1}) to be a Gaussian white noise. Under this assumption, 
he derived an approximate expression for the velocity correlation 
function $\langle v(t) v(0)\rangle$ 
which generalises the well-known, exponentially-decaying velocity 
correlation function of the linear 
Langevin equation with viscous friction \cite{gardiner1985}. He also obtained the 
stationary velocity distribution, previously reported
in \cite{kawarada2004,hayakawa2005}.
 
In this paper, we extend de~Gennes's work by calculating the 
time-dependent propagator $p(v,t|v_0,0)$ of the Langevin equation 
\be 
m\dot v=-\alpha v-\Delta_F \sign(v)+F+\xi(t), 
\label{eqd1} 
\ee 
which includes a viscous-like friction force with drag $\alpha$, 
a dry friction force with threshold force $\Delta_F$, a constant external 
force $F$, and a Gaussian white noise $\xi(t)$ characterised by 
\be
\langle \xi(t)\rangle=0,\qquad \langle \xi(t)\xi(0)\rangle=m^2 \Gamma \delta(t),
\ee 
with $m^2 \Gamma$  the noise power. In \cite{baule2010,baule2010b}, the propagator of 
(\ref{eqd1}) was obtained using path integral techniques in the small-noise limit,
$\Gamma\rightarrow 0$.  
Here we obtain the propagator without any approximation by directly 
solving the time-dependent 
Fokker-Planck equation associated with (\ref{eqd1}), which is 
explicitly given by 
\be
\frac{\partial p(v,t|v_0,0)}{\partial t}=\frac{\partial}{\partial v} 
(\gamma v + \Delta \sign(v)-a) p(v,t|v_0,0) 
+\frac{\Gamma}{2}\frac{\partial^2 p(v,t|v_0,0)}{\partial v^2},
\label{vfpl}  
\ee 
where $\gamma=\alpha/m$, $\Delta=\Delta_F/m$, and $a=F/m$ denote the 
force parameters rescaled by the mass $m$. To approach the solution
of the Fokker-Planck equation, we first consider in section \ref{secdf} the case
where only dry friction is present, i.e., $\gamma=F=0$, in order to establish the notation
and to restate the results of de~Gennes \cite{gennes2005}. 
The effects of the viscous force is then studied in
section \ref{secdvf}. In section \ref{secextf}, we finally
consider the case where all forces are present in order to
study the stick-slip transition described above, but now in the presence of noise. 
We shall see there that, although there is strictly speaking no sticking 
in the presence of Gaussian white noise \cite{gennes2005}, the Langevin dynamics
with dry friction does display two phases that resemble
the stick and slip states of the noiseless dynamics.  

Apart from extending de~Gennes's results to include viscous damping and
external forcing, the model studied here is interesting 
in part because it provides an example of piecewise-smooth
dynamical system perturbed by noise \cite{bernardo2008} which 
can be solved exactly by analytical means. 
In the case of pure dry friction, we are indeed able to obtain 
closed-form expressions for the time-dependent propagator
and the velocity correlation function, 
while for the dry-viscous and full-force cases, these functions are
 written in terms of spectral sums involving the so-called
parabolic cylinder function. These exact results are valid for any noise power, 
and so also extend those obtained in
\cite{baule2010,baule2010b}.

In a sense, it should not come as a surprise that the
Fokker-Planck equation (\ref{vfpl}) can be solve exactly, since it is one-dimensional and
piecewise linear. However, the solution is non-trivial and provides some 
physical insights into
stick-slip transitions perturbed by noise. In particular, we are able to identify from it
a number of time scales, expressed in terms of the noise power and force coefficients, 
which could be valuable inputs for experiments on driven systems with dry friction 
and driven droplet dynamics, such as those performed recently by 
Chaudhury et al.~\cite{daniel2002,chaudhury2008,goohpattader2009,goohpattader2010}.
Our results on the spectrum of the Fokker-Planck operator of (\ref{vfpl}) also enable us to 
state conditions under which the viscous force can be neglected with
respect to the dry friction, a regime referred to by de Gennes \cite{gennes2005} as the partly stuck regime, 
or, conversely, under which dry friction can be neglected with respect to viscous friction (viscous regime). 
These conditions might be useful for determining whether dry friction 
or friction-like forces, such as liquid hysteresis \cite{daniel2002,chaudhury2008},
should be considered when modeling noise-driven systems.
 
\section{Dry friction only} 
\label{secdf} 
 
We recall in this section the solution of the Fokker-Planck equation (\ref{vfpl}) 
in the simple case where there is no viscous damping and no external force, i.e., 
$\gamma=0$, $a=0$, $\Delta>0$ and $\Gamma>0$, for the purpose of introducing
the general notation used throughout the paper and recalling the results
of de~Gennes \cite{gennes2005}.

The solution proceeds in the standard way 
(see, e.g., \cite{Risk:89} and \cite{HoLe:84} for a rigorous but accessible account).
Using the appropriate nondimensional variables 
\be\label{ba} 
x=\frac{2 \Delta}{\Gamma} v, \quad \tau=\frac{2 \Delta^2}{\Gamma} t,
\ee 
we first recast the Fokker-Planck equation in the form 
\be
\frac{\partial p(x,\tau|x',0)}{\partial \tau}= \frac{\partial \Phi'(x) p(x,\tau|x',0)}{\partial x} +  
\frac{\partial^2 p(x,\tau|x',0)}{\partial x^2},
\label{aa} 
\ee 
where $x'$ stands for the initial condition, and $\Phi(x)=|x|$ is the nondimensional 
potential associated with the dry friction force. The normalised stationary solution of (\ref{aa}) is 
given in terms of this potential by 
\be\label{ab} 
\rho_*(x)=\frac{e^{-\Phi(x)}}{Z_*},
\ee 
where $Z_*$ is the normalisation factor. 

To find the time-dependent 
state $p(x,\tau|x',0)$, we next solve the eigenvalue equation 
\be\label{ac} 
-\Lambda u_\Lambda(x)=\left(\Phi'(x) u_\Lambda(x)\right)' + u''_\Lambda(x),
\ee 
for $\Lambda\ge 0$. The eigenfunctions $v_\Lambda$ of the adjoint problem are 
related to the eigenfunctions $u_\Lambda$ via the stationary distribution 
\be\label{ad} 
u_\Lambda(x) = v_\Lambda(x)\, e^{-\Phi(x)}. 
\ee 
The two sets of eigenfunctions form a bi-orthogonal system, so that the propagator can be  
formally written as 
\be\label{ae} 
p(x,\tau|x',0) = \sum_\Lambda e^{-\Lambda \tau} u_\Lambda(x) v_\Lambda(x')/Z_\Lambda 
\ee 
with normalisation 
\be\label{af} 
Z_\Lambda = \int_{-\infty}^\infty v_\Lambda(x)\, u_\Lambda(x)\, dx.
\ee 

Since we are concerned with a piecewise smooth potential whose derivative, representing the force, 
shows a finite jump  at the origin, we have to solve (\ref{ac}) on the positive 
and negative domains separately,
with standard, decaying boundary conditions at infinity, and then match at the origin the 
two solutions obtained by imposing 
continuity of the eigenfunction and continuity of the probability current
\be
j(x)=\Phi'(x) u_\Lambda(x)+u'_\Lambda(x).
\ee 
This yields the conditions
\begin{eqnarray}
u_\Lambda(0-) &=& u_\Lambda(0+) \nonumber \\ 
\Phi'(0-) u_\Lambda(0-)+u'_\Lambda(0-)&=& 
\Phi'(0+) u_\Lambda(0+)+u'_\Lambda(0+). 
\label{agb} 
\end{eqnarray} 
Additionally, we impose the initial condition $p(x,0|x',0)=\delta(x-x')$.

The complete analytical solution of (\ref{ac}) with the above boundary conditions can be easily written 
down; see, e.g., \cite{Wong_AMS64} for a classical account. To keep the presentation 
self-contained, we summarise
in Appendix~\ref{secwed} the major steps leading to the final solution 
given by
\be
p(x,\tau|x',0) = \frac{e^{-\tau/4}}{2 \sqrt{\pi \tau}} e^{-(|x|-|x'|)/2}  
e^{-(x-x')^2/(4\tau)} + \frac{e^{-|x|}}{4}  
\left[1 + \erf\left(\frac{\tau-(|x|+|x'|)}{2\sqrt{\tau}}\right) \right],
\label{bb} 
\ee
where $\erf(x)=2 \int_0^x \exp(-z^2) dz/\sqrt{\pi}$ denotes the probability integral. 

\begin{figure}[t] 
\centerline{\includegraphics[width=0.45\linewidth]{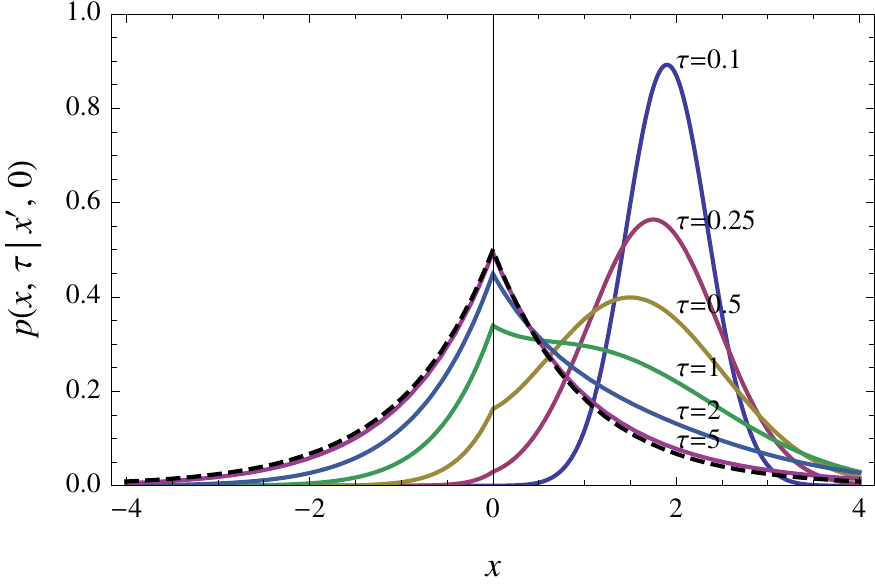}} 
\caption{(Color online) Propagator $p(x,\tau|x',0)$  
of the Fokker-Planck equation for the pure dry friction case 
with initial condition $x'=2$. The dashed line shows the stationary solution.} 
\label{figproppure1} 
\end{figure} 

Figure \ref{figproppure1} shows the behaviour of this solution as $\tau$ 
increases. At a qualitative level, the behaviour of $p(x,\tau|x',0)$ shows two stages: a short time relaxation of the
initial distribution towards the origin with the build-up of a corner or cusp at the origin, 
followed, on a longer time scale, by a build-up of
the exponential tails of the stationary state $\rho_*(x)$. The first stage 
can be understood superficially by noting that the deterministic (zero-noise) dynamics of the model 
admits the solution $v(t)=v_0-\Delta t$ with $v_0>0$,
and thus shows a finite-time relaxation towards the fixed point $v=0$ after a time
$t_c=v_0/\Delta$. In the nondimensional variables of (\ref{ba}), this time-scale reads
$\tau_c=x'$ and is the time-scale within which the propagator develops its
unimodal, exponential-type shape (see figure \ref{figproppure1}).

The effect of the noise on the finite-time decay of $v(t)$ can be seen by computing the expectation of $x(\tau)$, i.e., the first moment of (\ref{bb}):
\be
\label{bba}
\langle x(\tau)\rangle =\sigma(x') \left[ (|x'|-\tau) 
\frac{\erfc(-(|x'|-\tau)/(2 \sqrt{\tau}))}{2} \right. 
+ \left.(|x'|+\tau)\frac{\exp(|x'|)\, \erfc\left((|x'|+\tau)/(2\sqrt{\tau})\right)}{2}
\right].
\ee
Here $\erfc(x)=1-\erf(x)$ denotes the complementary error function.
If one considers cases with large initial conditions ($|x'|\gg 1$), i.e.,
cases where the noise can be considered as a small perturbation, then the
initial stage $\tau <|x'|$ is dominated by the first contribution to
(\ref{bba}), and the mean value follows the deterministic path
$x'-\tau \sigma(x')$, apart from exponentially small corrections. Beyond
this time scale, i.e., $\tau>|x'|$, both contributions are of the same size, and the
mean value follows an exponential relaxation towards its stationary value $x_*=0$,
dominated by diffusive behaviour.

To complement this analysis, we can compute the correlation function, defined by
\be
\langle x(\tau) x(0) \rangle = \int_{-\infty}^\infty dx \int_{-\infty}^\infty dx'\, x\, x'\, p(x,\tau| x',0) \rho_*(x').
\ee
Because of the spectral gap observed in the spectrum of the Fokker-Planck operator
(see Appendix~\ref{secwed}), we know that $\langle x(\tau) x(0) \rangle$ must decay exponentially for large times;
however, because of the finite-time decay of the dry friction dynamics, we expect to see corrections to 
this exponential decay for short times. This can be verified explicitly by computing the two integrals 
appearing in the expression of $\langle x(\tau) x(0) \rangle$ and by noting that
the symmetric part of the propagator (\ref{bb}) does not contribute in these integrals. The result is
\begin{eqnarray}\label{bc} 
\langle x(\tau) x(0) \rangle &=& \frac{e^{-\tau/4}}{\sqrt{\pi \tau}}  \int_0^\infty dx \int_0^\infty dy\,
xy\, e^{-(x+y)/2}\, e^{-(x^2+y^2)/(4\tau)} \sinh( xy/(2\tau))\nonumber \\ 
&=&  \frac{e^{-\tau/4}}{6 \sqrt{\pi\tau}} \left\{ \left( \frac{\sqrt{\pi\tau}}{2}  e^{\tau/4} 
\erfc(\sqrt{\tau}/2)-1 \right) \left(\tau^3+6\tau^2-12 \tau +24\right) + 2\tau^2+24 \right\} 
\nonumber \\ 
&=& \frac{16}{\sqrt{\pi}} \frac{e^{-\tau/4}}{(\tau/4)^{3/2}}[1+{\cal O}(\tau^{-1})],
\end{eqnarray} 
where, in the last step, the standard asymptotic property of the probability integral 
was used. As usual, we can define a correlation time $\tau_\corr$ from this result by putting the dominant
exponential term in the form $e^{-\tau/\tau_\corr}$ to obtain $\tau_\corr=4$ or
\be
\label{corrt}
t_\corr=\frac{2m^2\Gamma}{\Delta_F^2}
\ee
in the original variables given in (\ref{ba}). This correlation time and the last line of (\ref{bc})
recover the results of de~Gennes \cite{gennes2005} up to some constants.\footnote{There is a typo in equation~(37) 
of \cite{gennes2005}: $\tau\Delta$ should be $\tau_\Delta$. There also seems to be
a missing factor 2 in some of de Gennes's results.}

\section{Dry and viscous friction} 
\label{secdvf} 
 
For the case with viscous friction, $\gamma>0$, but without external force, $a=0$, the appropriate 
nondimensional units are given by 
\be\label{da} 
x=\left(\frac{2 \gamma}{\Gamma}\right)^{1/2} v , \qquad \tau=\gamma t.
\ee 
In terms of these variables, the Fokker-Planck equation (\ref{vfpl}) takes the form (\ref{aa}) 
with the potential 
\be\label{db} 
\Phi(x)=(|x|+\delta)^2/2, \qquad \delta= \Delta 
\left(\frac{2}{\gamma \Gamma}\right)^{1/2}. 
\ee 
The parameter $\delta$ measures the strength of the dry friction relative to the 
viscous damping and is therefore positive. However, a negative $\delta$ is possible if we have in mind 
to study Kramer's classical problem of a Brownian particle diffusing in a bistable or double-well potential. 
We will mention this case later in the paper.

Since the potential (\ref{db}) is piecewise parabolic, 
the eigenvalue problem of the Fokker-Planck operator can still be solved analytically.
However, to the best of our knowledge, a closed analytical expression of the propagator 
is not known. Various details can be found in the literature, in particular,
in the context of the exit-time problem for the Ornstein-Uhlenbeck process 
\cite{Sieg_PR51}
and the constrained harmonic quantum oscillator \cite{Dean_PCPS66,auluck1945}.
As a coherent presentation cannot easily be found, we provide some details in 
Appendix~\ref{secsym}. 
 
\begin{figure}[t] 
\centerline{\includegraphics[width=0.9\linewidth]{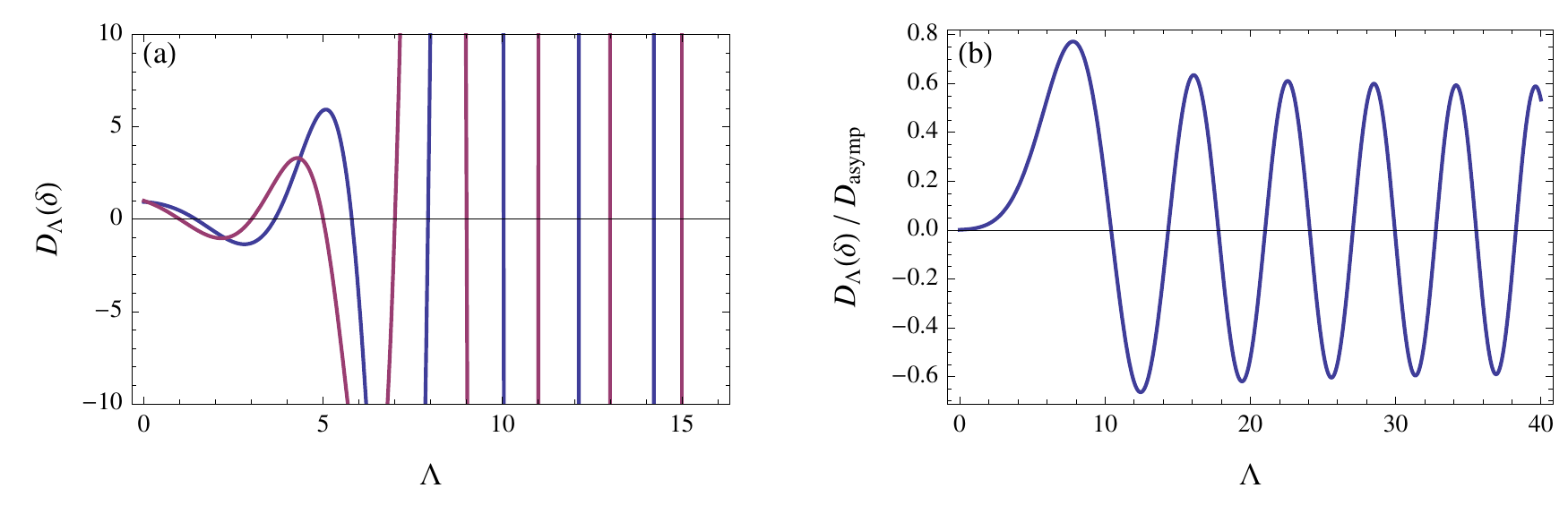}}
\caption{(Color online) (a) Parabolic cylinder function $D_\Lambda(\delta)$ as a function of the 
index $\Lambda$ for $\delta=0$ (purple) and $\delta=0.5$ (blue). The zeros yield the eigenvalues 
$\Lambda^{(o)}_n$ which, for $\delta=0$, are the odd positive integers. 
(b) Parabolic cylinder function for $\delta=5$ normalised by the asymptotic growth 
$D_{\rm asymp}=2^{\Lambda/2} \Gamma ((1+\Lambda)/2)$ for large $\Lambda$; 
see \cite{AbSt:70}. The resulting expression remains bounded.
\label{fig1}}  
\end{figure}

Because of the quadratic shape of the potential (\ref{db}), we expect the 
spectrum to be pure point.\footnote{Elliot's Theorem \cite{HoLe:84} 
does not apply here, so a rigorous proof of the pure point property cannot easily be given.}
The eigenfunctions are either even or odd because of the symmetry of the potential. 
The eigenvalues $\Lambda_n^{(o)}$ associated with the odd eigenfunctions  
are determined for $n\geq 1$ by the characteristic equation
\be\label{dc} 
D_{\Lambda^{(o)}_n}(\delta)=0,
\ee 
where $D_\Lambda(z)$ denotes the parabolic cylinder function 
\cite{AbSt:70} (see figure \ref{fig1}).\footnote{As expected, all the
eigenvalues are positive, since the parabolic cylinder function is strictly
positive for negative values of the index $\Lambda$.}
Apart from the stationary solution with eigenvalue $\Lambda_0=0$, the other 
eigenvalues associated with the even eigenfunctions are simply given by the relation 
\be\label{dd} 
\Lambda^{(e)}_n=\Lambda^{(o)}_n +1. 
\ee 

\begin{figure}[t] 
\centerline{\includegraphics[width=0.45\linewidth]{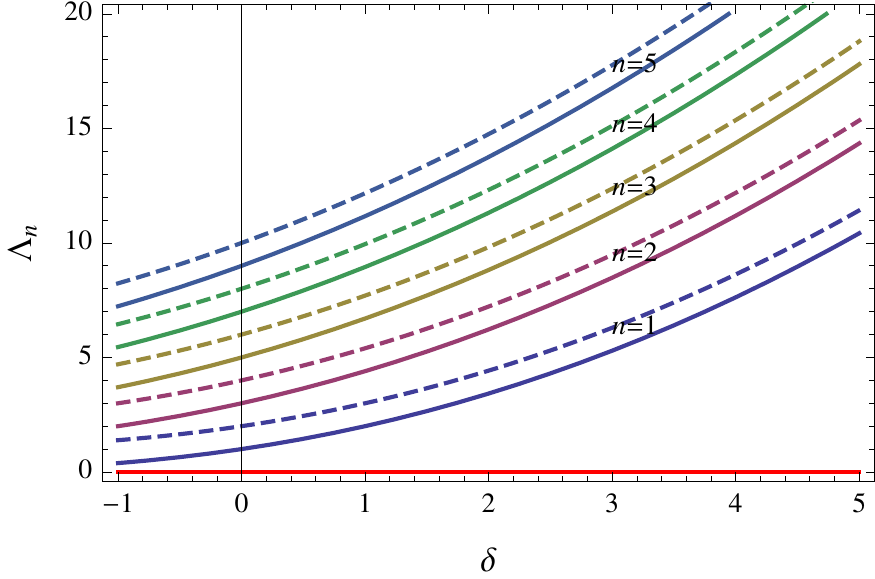}}
\caption{(Color online) Eigenvalues of the dry friction and viscous 
friction problem as a 
function of $\delta$; see (\ref{dc}) and (\ref{dd}). The red 
line corresponds to
$\Lambda_0=0$, while the solid and dashed lines correspond, respectively, 
to the odd and even eigenvalues for $n=1,2,3,4$, and $5$ (see also
\cite{Dean_PCPS66,auluck1945,RiSh_JAP88}).
\label{fig1a}}  
\end{figure}

The case without dry friction, i.e., $\delta=0$, results in a quadratic potential 
associated with the Ornstein-Uhlenbeck process. Even and odd eigenvalues are then given by positive 
even and odd integers, as the characteristic equation (\ref{dc}) reduces to an equation involving
Hermite polynomials. For negative values of $\delta$, the potential becomes bistable and 
the lowest non-trivial eigenvalue approaches zero, reflecting the behaviour of Kramer's 
escape rate \cite{Risk:89}. For large positive values of $\delta$, the spectrum develops
a gap between the two lowest eigenvalues, $\Lambda_0$ and $\Lambda_1^{(o)}$, 
and the rest of the spectrum. These properties of the characteristic equation and of the spectrum are 
illustrated in figures \ref{fig1} and \ref{fig1a}. 

\begin{figure}[t]
\centerline{\includegraphics[width=0.9\linewidth]{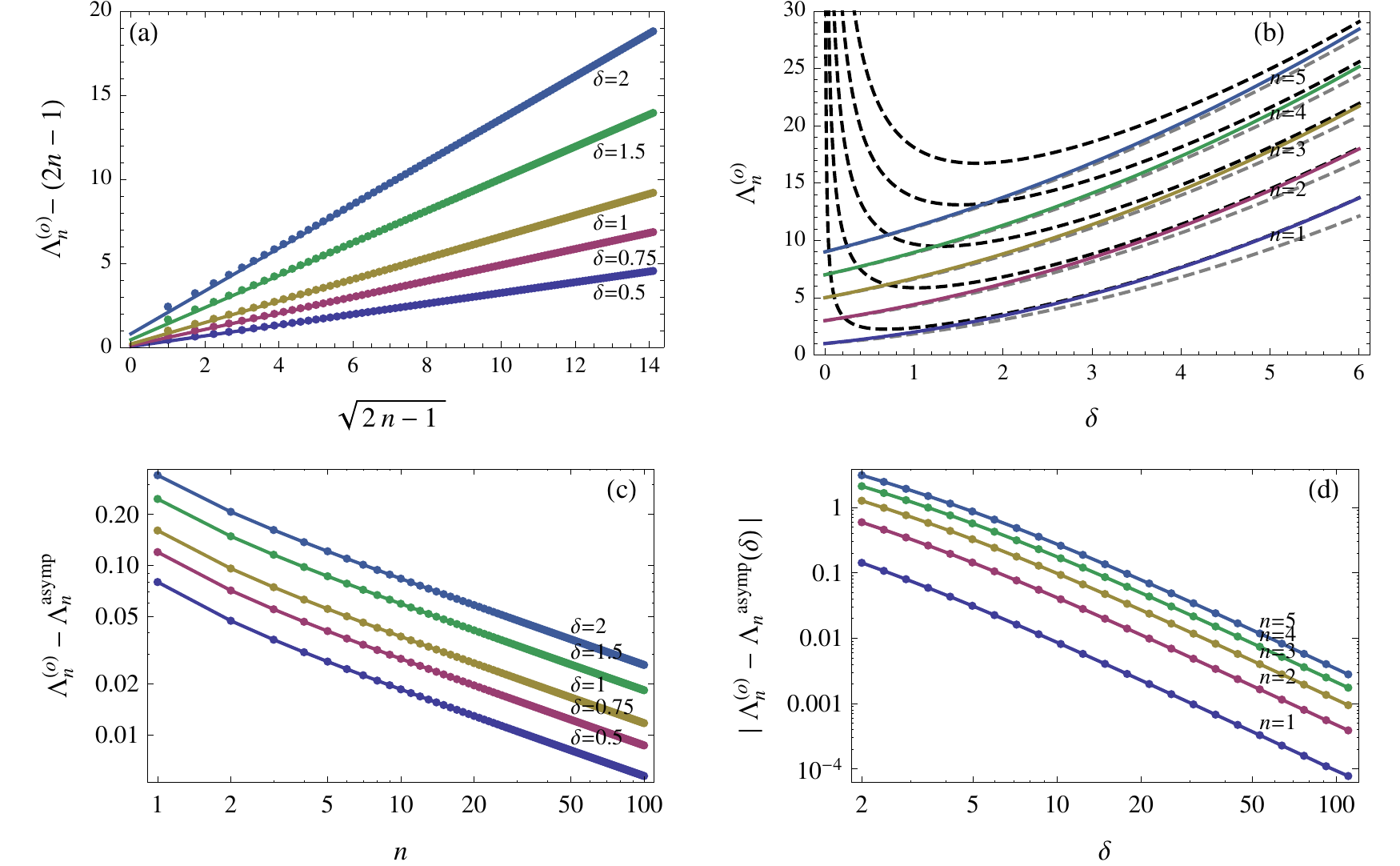}}
\caption{(Color online) Eigenvalues for the dry and viscous friction case. (a) Illustration of 
the asymptotic relation (lines)
satisfied by $\Lambda_n^{(o)}$ (symbols) for large index $n$; see 
(\ref{dda}).
(b) $\Lambda_n^{(o)}$ (see also figure \ref{fig1}) 
as a function of $\delta$ together with the asymptotic expressions 
(\ref{dda}) (gray dashed line) 
and (\ref{ddb}) (black dashed line). A crossover between the small $\delta$ (large $n$) 
and large $\delta$ (small $n$) asymptotics is clearly visible.
(c) Log-log plot of the difference between the eigenvalue $\Lambda_n^{(o)}$ and
the asymptotic expression (\ref{dda}), referred to as $\Lambda_n^{\rm asymp}$ in the plot,
as a function of $n$. The slope of the observed straight lines is approximately $-1/2$.
(d) Log-log plot of the distance between the eigenvalue $\Lambda_n^{(o)}$ and the asymptotic expression of (\ref{ddb}),
referred to as $\Lambda_n^{\rm asymp}(\delta)$ in the plot, as a function of $\delta$.
The slope of the observed straight lines is approximately $-2$.} 
\label{figeigenvalvisc1} 
\end{figure} 
 
Given the closed analytic expression of the characteristic equation, we can 
use standard asymptotic expansions for the parabolic cylinder function 
to obtain 
estimates for the eigenvalues \cite{AbSt:70}. 
In this respect, the expansion of 
the oscillating part of the parabolic cylinder function, e.g., in terms of Airy functions, is 
quite useful, as it enables us to obtain expressions for the eigenvalues for large index $n$ 
and for large values of the parameter $\delta$. For the upper part of the spectrum, we explicitly obtain
\be\label{dda} 
\Lambda_n^{(o)} \simeq (2n-1) +\frac{2 \delta}{\pi} \sqrt{2n-1} +2 \left(\frac{\delta}{\pi}\right)^2,\qquad 
n \gg 1. 
\ee 
This asymptotic relation is illustrated in figure 
\ref{figeigenvalvisc1}(a). From figure \ref{figeigenvalvisc1}(c), we
see that the error associated with (\ref{dda}) is of order $\mathcal{O}(n^{-1/2})$ in $n$. It should not come as a
surprise that the leading term in (\ref{dda}) coincides with the 
spectrum of the Ornstein-Uhlenbeck process. However, it is not entirely obvious that the  
leading correction should also increase with $n$. Such a feature is explained by noting that the
mismatch in width between the piecewise parabolic potential 
$\Phi(x)$ and the fully parabolic potential of the Ornstein-Uhlenbeck process
increases in size when $x$ increases. 

For large values of the parameter $\delta$ and fixed 
values of $n$, the asymptotic property of the parabolic cylinder function results in 
\be\label{ddb} 
\Lambda_n^{(o)}+1/2 \simeq \frac{\delta^2}{4} \left[ 1 +  
t_n \left(\frac{\delta^2}{4}\right)^{-2/3} + \frac{2 t_n^2 }{15}  \left(\frac{\delta^2}{4}\right)^{-4/3} 
\right], \qquad \delta \gg 1,
\ee 
where $-t_n$ labels the zeros of the Airy function $\mbox{Ai}(z)$. This expression, which is illustrated
in figure \ref{figeigenvalvisc1}(b), captures the 
limiting case of vanishing viscous friction and the transition from a discrete to a continuous spectrum. One clearly 
observes the spectral gap of size $\delta^2/4$, and the fact that the eigenvalues accumulate for 
large $\delta$ in a quasicontinuous way when rescaled by $\delta^2$ to account for the different time scales defined in
(\ref{ba}) and (\ref{da}). 

From figure \ref{figeigenvalvisc1}(d),
we are led to conjecture that the error term associated with (\ref{ddb}) is of order $\mathcal{O}(\delta^{-2})$.
Note also that if we allow for negative $\delta$, then 
the asymptotic expansion of the parabolic cylinder function yields, as expected, 
an activation rate result for the Kramer escape problem, having the form 
\be\label{ddc} 
\Lambda_{1}^{(o)} \simeq \frac{-\delta\, e^{-\delta^2/2}}{\sqrt{2 \pi} A(\delta)}, \qquad -\delta \gg 1,
\ee 
with the denominator of the pre-exponential factor being given by the asymptotic  
series\footnote{This expression determines a nonnegative bounded real function.} 
\be\label{ddd} 
A(\delta) = \sum_{k=0}^\infty \frac{(2k)!}{k!} \left(\frac{1}{2 \delta^2}\right)^k \simeq 
-\sqrt{\pi} e^{-\delta^2/2} \frac{i \delta}{\sqrt{2}}\, \erf\left(\frac{i \delta}{\sqrt{2}}\right). 
\ee 
 
Having studied the spectrum of the Fokker-Planck operator in detail, we now turn to the propagator.
As in the pure dry friction case, the eigenfunctions can
be classified as odd or even and are now given
in terms of parabolic cylinder functions:
\be\label{de} 
u^{(e/o)}_n(x) = \exp(-(x+\delta)^2/4)\, D_{\Lambda^{(e/o)}_n}(x+\delta), \qquad x\geq 0, 
\ee 
where the expression for negative $x$ follows by symmetry. The eigenfunctions  
$v^{(e/o)}_n(x)$ of the adjoint problem are, as usual, determined by 
(\ref{ad}). 
The normalisation of the odd eigenfunctions reads\footnote{$\partial_\Lambda
D_\Lambda(\delta)$ stands for the derivative of the parabolic cylinder function with
respect to its index. In addition to the standard derivation given in Appendix~\ref{secsym}, 
the value of the normalisation integral can be computed by comparing 
the exact result of the Laplace transform of the exit time distribution 
\cite{Sieg_PR51} with 
the Laplace transform of the odd part of the propagator shown in (\ref{dh}).}  
\be
\label{df} 
Z_n^{(o)} = \int_{-\infty}^{\infty} v_n^{(o)}(x) u_n^{(o)}(x)\, dx = 2 \Lambda_n^{(o)} D_{\Lambda_n^{(o)}-1}(\delta) 
\left(-\partial_{\Lambda_n^{(o)}} D_{\Lambda_n^{(o)}}(\delta) \right),
\ee 
and yields the normalisation of the even eigenfunctions via 
\be\label{dg} 
Z_n^{(e)} = \int_{-\infty}^{\infty} v_n^{(e)}(x) u_n^{(e)}(x)\, dx = (\Lambda_n^{(o)}+1) Z_n^{(o)}. 
\ee 
From these expressions, we obtain the spectral representation of 
the propagator as
\be
\label{dh} 
 p(x,\tau|x',0) = \rho_*(x)+ \sum_{n=1}^\infty \exp(-\Lambda_n^{(o)} \tau) 
\frac{u_n^{(o)}(x) v_n^{(o)}(x')}{Z_n^{(o)}} 
+ \sum_{n=1}^\infty \exp(-\Lambda_n^{(e)} \tau) 
\frac{u_n^{(e)}(x) v_n^{(e)}(x')}{Z_n^{(e)}}. 
\ee
More details about this result can be found in Appendix~\ref{secsym}.

\begin{figure}[t] 
\centerline{\includegraphics[width=0.45\linewidth]{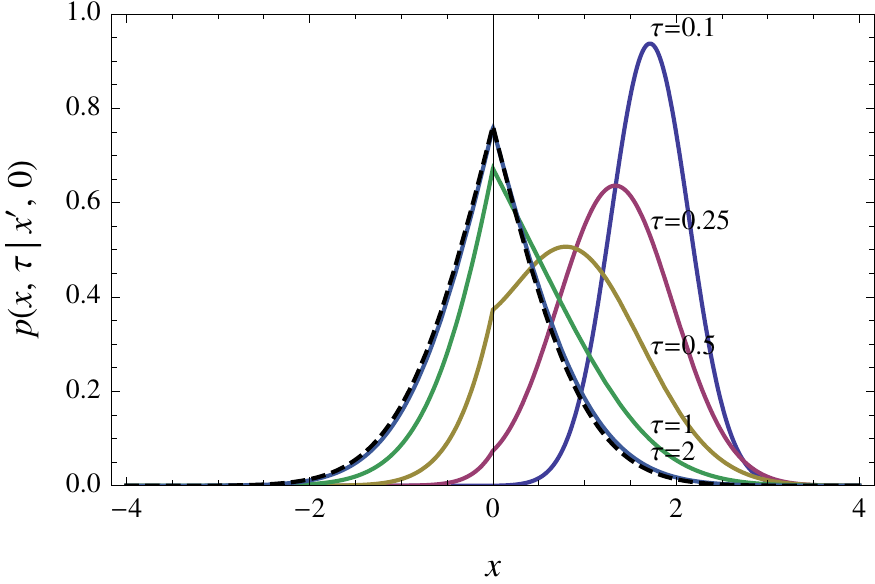}} 
\caption{(Color online) Dry friction with viscous friction. Evolution of $p(x,\tau|x',0)$
 towards the stationary distribution (dashed line) as $\tau$ increases. Parameters: $x'=2$
 and $\delta=1$. 40 modes were used to plot the propagator for the two first times; for the
 the later times, only 5 modes were used.} 
\label{figpropvisc1} 
\end{figure} 

The behaviour of the propagator (\ref{dh}) for finite values of $\delta$ 
is illustrated in figure \ref{figpropvisc1}. Overall, we see that
for moderate values of the viscous damping, i.e., for moderate positive values of the parameter
$\delta$, the evolution of $p(x,\tau|x',0)$ largely resembles the case of pure dry friction with its cusp at the origin; 
see figure \ref{figproppure1}. The only minor difference is that $p(x,\tau|x',0)$ converges 
to the stationary distribution $\rho_*(x)$ relatively faster than with pure dry friction 
because of the added viscous friction. This property is reflected in the value of the lowest non-trivial eigenvalue
at large $\delta$; see equation (\ref{ddb}).

In the absence of dry friction ($\delta=0$), the potential $\Phi(x)$ becomes quadratic
and the problem reduces to the Ornstein-Uhlenbeck process, as mentioned before. In this case, the parabolic cylinder functions for 
integer indices can be expressed in terms of Hermite polynomials
\be
D_n(z)= e^{-z^2/2}\, \mbox{He}_n(z),
\ee
and lead to odd and even eigenvalues corresponding to odd and even positive integers, respectively (see figure~\ref{fig1a}). In this special case, the spectral  
decomposition (\ref{dh}) can be written in closed analytical form using Mehler's formula \cite{HoLe:84}. 

\begin{figure}[t] 
\centerline{\includegraphics[width=0.9\linewidth]{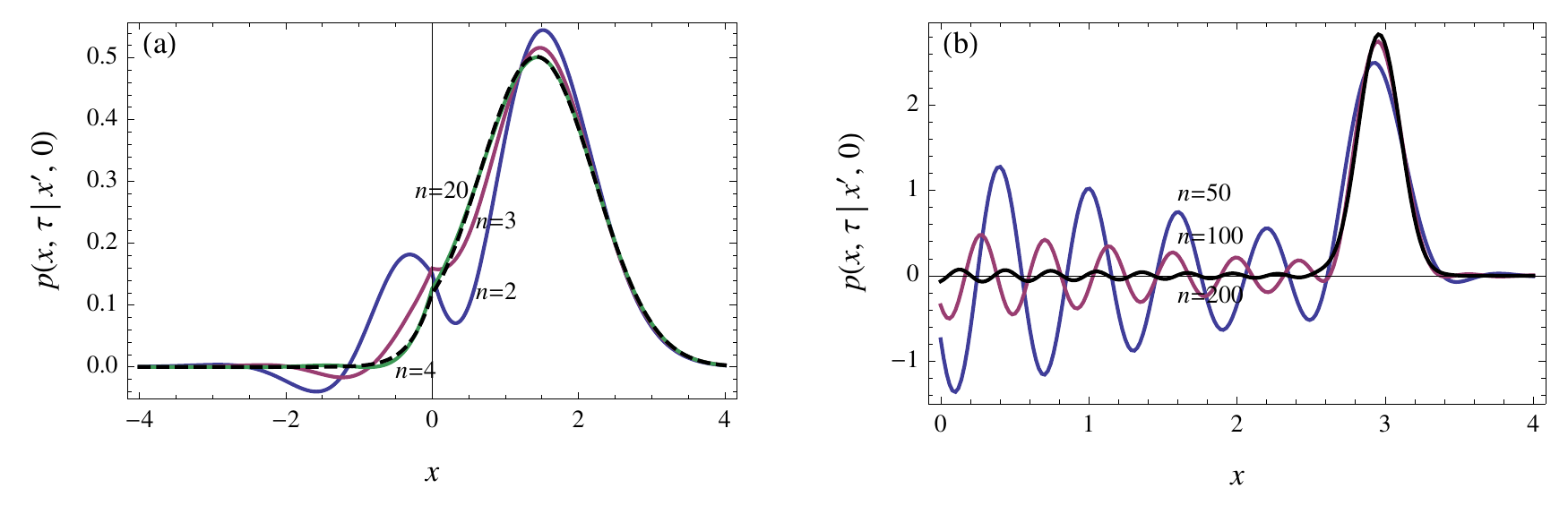}}
\caption{(Color online) Dry friction with viscous friction. (a) Convergence of the mode 
decomposition of $p(x,\tau|x',0)$ for $n=2,3,4$ modes (coloured lines) and $n=20$ modes (dashed black line). 
Parameters: $x'=3$, $\delta=1$ and $\tau=0.5$. (b) Mode convergence for short time 
with $n=10,50,100,200$ modes. Parameters: $x'=3$, $\delta=1$ and $\tau=0.01$.} 
\label{figpropmodes1} 
\end{figure}

For the general case where dry and viscous frictions are present, 
we are not aware of a closed analytical expression for the propagator of 
(\ref{dh}).
Thus, to evaluate this propagator, and to generate, for instance, the curves shown in 
figure \ref{figpropvisc1}, we have to rely on the numerical computation of the spectral sum. 
The behaviour of this sum as the number of eigenfunctions or modes is increased is shown 
in figure \ref{figpropmodes1}. As one expects from the result
of (\ref{dh}), one requires a large number of modes to capture the short time behaviour 
of $p(x,\tau|x',0)$,
and a Gibbs-type phenomenon is visible when an insufficient number of modes is used.
This problem becomes irrelevant when the dynamics on larger time scales is considered, 
as modes in the upper part of the spectrum are strongly damped.
  
For the correlation function, only the odd part of the propagator (\ref{dh}) contributes. 
Using (\ref{ad}) for the adjoint eigenfunctions and (\ref{ab}) for the 
stationary distribution, we obtain the representation 
\be\label{di} 
\langle x(\tau) x(0) \rangle =  \sum_{n=1}^\infty  
\frac {\exp(-\Lambda_n^{(o)} \tau)}{Z_n^{(o)}} 
\frac{2 \left( \int_0^\infty x\, u_n^{(o)}(x)\, dx \right)^2} 
{\int_0^\infty \exp(-(x+\delta)^2/2)\, dx}. 
\ee 
The numerator of this expression can be simplified using a recurrence relation for the parabolic  
cylinder functions reproduced in (\ref{efa}) and (\ref{efb}), the representation (\ref{de}) 
for the eigenfunctions, and integration by parts
to obtain
\be\label{dj} 
\int_0^\infty x\, u_n^{(o)}(x)\, dx = e^{-\delta^2/4}\, D_{\Lambda_n^{(o)}-2}(\delta). 
\ee 
Thus, the series (\ref{di}) reads
\begin{eqnarray}
\label{dja} 
\langle x(\tau) x(0) \rangle &=& \frac4{\sqrt{2\pi}}\frac{\exp(-\delta^2/2)}{\erfc(\delta/\sqrt{2})}  
\sum_{n=1}^\infty  
\frac{\exp(-\Lambda_n^{(o)} \tau)}{Z_n^{(o)}}D^2_{\Lambda_n^{(o)}-2}(\delta) \nonumber \\ 
&=& \frac{\delta^2}{\sqrt{\pi/2} \exp(\delta^2/2)\erfc(\delta/\sqrt{2})} 
\sum_{n=1}^\infty \frac{\exp(-\Lambda_n^{(o)}\tau)}{\Lambda_n^{(o)}(\Lambda_n^{(o)}-1)^2} 
\frac{D_{\Lambda_n^{(o)}-1}(\delta)}{(-
\partial_{\Lambda_n^{(o)}} D_{\Lambda_n^{(o)}}(\delta))},
\end{eqnarray} 
where for the last step we have used (\ref{df}) and (\ref{dc}), 
as well as the recurrence relation (\ref{djb}).

\begin{figure}[t] 
\centerline{\includegraphics[width=0.45\linewidth]{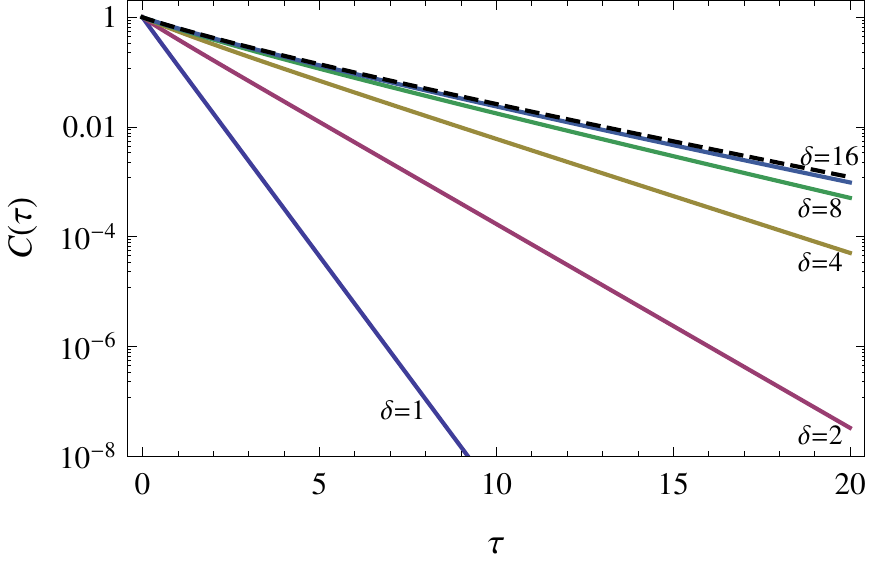}}
\caption{(Color online) Dry friction with viscous friction. Log-linear plot of the 
normalised correlation function 
$C(\tau)=\langle x(\tau) x(0) \rangle/\langle x^2\rangle_*$ 
for $\delta=1$, $2$, $4$, $8$, and $16$; see (\ref{dja}).
The dashed line corresponds to the case without viscous damping, 
obtained in the limit $\delta \rightarrow \infty$; see (\ref{bc}).
For the purpose of comparing the dry and viscous friction case with the pure dry friction case, 
the data are displayed in the units defined in (\ref{ba}).}
\label{corr1} 
\end{figure} 

Figure \ref{corr1} confirms that the correlation function (\ref{dja}) decays exponentially. 
This agrees with the spectral gap of the spectrum of 
the Fokker-Planck operator, which, according to equation (\ref{ddb}), has a size of order $\delta^2/4$.
For increasing values of $\delta$, the correlation function of (\ref{dja}) approaches the pure dry friction
result of de~Gennes \cite{gennes2005} shown in (\ref{bc}). 
In fact, for $\delta \sim 10$, the correlation function in the presence of dry and viscous frictions 
is almost indistinguishable from that obtained with dry friction only.
From this, we conclude that a viscous damping of order $\gamma \Gamma \sim 0.01 \Delta^2$
does not affect de~Gennes's result significantly. Since this result
is based on the product $\gamma \Gamma$ being small, we also conclude that de~Gennes's result
is a good approximation of the correlation function for the dry and viscous friction case
when the noise is sufficiently weak.

The effect of viscous damping becomes apparent only at large times, as a result of the discrete spectrum 
obtained with dry and viscous friction. The asymptotic result of (\ref{ddb})
shows that the level spacing of the eigenvalues is of order $\delta^{2/3}$, so that at a time
scale of order $\tau\sim \delta^{-2/3}$, we start to see a difference between the exponential decay of
$\langle x(\tau) x(0)\rangle$
obtained with and without viscous damping. 
In the original units, this time scale is approximately given by
\be
t \sim t_\corr/(\gamma\, t_\corr)^{2/3},
\ee
where $t_\corr$ is the correlation time of the pure dry friction case, defined before in (\ref{corrt}).
The influence of
viscous damping on the tail of the correlation function is thus visible only if $\gamma t_\corr$ is not too large. 
One can study this regime in more detail by using asymptotic expansions for the parabolic cylinder 
function in the result of (\ref{dja}).

To conclude this section, note that in the absence of dry friction, $\delta=0$, 
the sum in (\ref{dja}) contains only 
one mode, since the numerator $D_{\Lambda_n^{(o)}-1}(0)$ vanishes for $n\geq 2$
because of the characteristic equation. This mode recovers, as expected, 
the correlation function of the
Ornstein-Uhlenbeck process with viscous friction.
  
\section{Constant external force} 
\label{secextf} 
 
\begin{figure}[t]
\centerline{\includegraphics[width=0.9\linewidth]{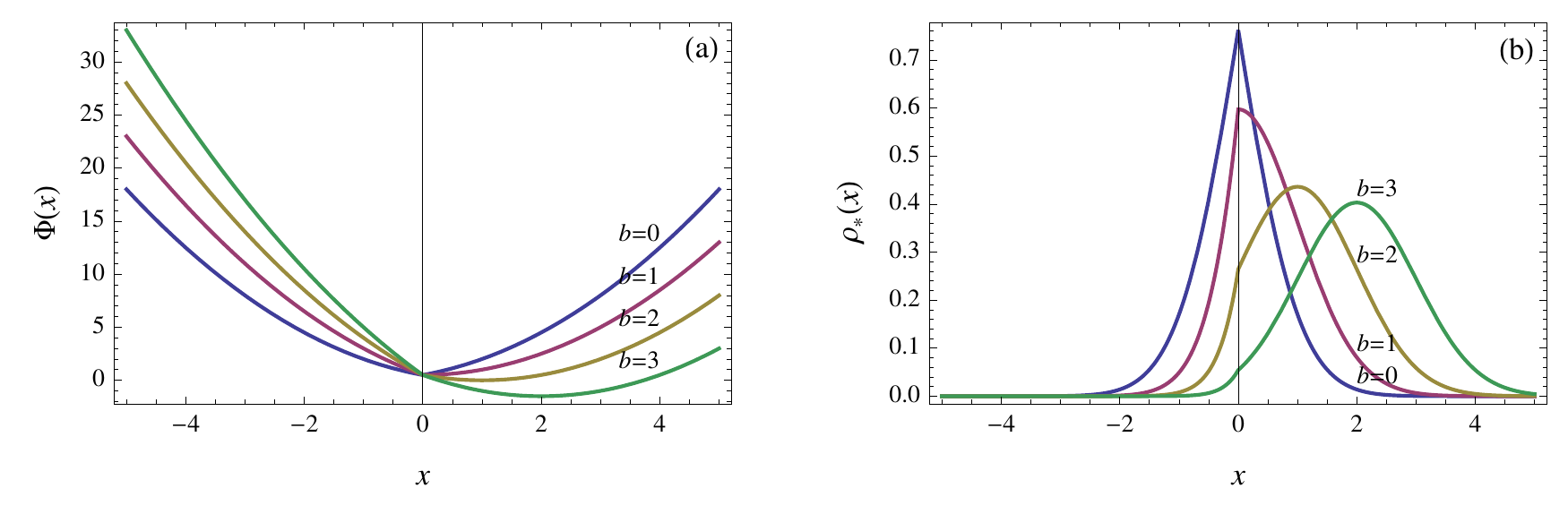}}
\caption{(Color online) Dry friction with viscous friction and forcing. 
(a) Potential $\Phi(x)$ as a function of $x$ for $\delta=1$ and for different values of $b$, as given by (\ref{fa}). 
(b) Corresponding invariant distribution, given by (\ref{ab}).}
\label{figinvdistforced1}
\end{figure}

For a system with constant external force, the suitable nondimensional 
variables are again given by (\ref{da}), and the potential $\Phi(x)$ which determines the drift 
of the Fokker-Planck equation reads 
\be\label{fa} 
\Phi(x)=\frac{(|x|+\delta)^2}{2}-b x, \quad \delta=\Delta \left(\frac{2}{\gamma\Gamma}\right)^{1/2},  
\quad b=a  \left(\frac{2}{\gamma\Gamma}\right)^{1/2}. 
\ee 
The additional parameter $b$ determines the strength of the ``acceleration'' $a$ relative to the 
viscous damping. Depending on the relative size of the two parameters 
$\delta$ and $b$, the deterministic part of the Langevin equation (\ref{eqd1}) 
either corresponds to a particle at rest or to a moving particle, as mentioned in the introduction.
For $|b|<\delta$, the deterministic steady state corresponding 
to the minimum of the potential (\ref{fa}) has vanishing velocity (stick state), while for 
$|b|>\delta$, a steady state with finite velocity occurs (slip state). 
In the presence of noise, these stick and slip steady states become the most probable 
stationary states of the Langevin dynamics, and so determine the maximum of the stationary 
distribution $\rho_*(x)$. This is illustrated in figure \ref{figinvdistforced1}.

To solve the time-dependent Fokker-Planck equation for the full-force case,
we proceed along the lines of the previous sections. The main steps are given in Appendix~\ref{secbia};
here we only summarise the main properties of the spectrum of the Fokker-Planck operator, 
the propagator $p(x,\tau|x',0)$ and the corresponding correlation function. 

Unlike the previous case where $b=0$, eigenfunctions cannot be classified according 
to their symmetry when $b\neq 0$. The characteristic equation is given by 
\be\label{fb} 
\Lambda_n \left(D_{\Lambda_n}(\delta+b) D_{\Lambda_n-1}(\delta-b) 
+ D_{\Lambda_n}(\delta-b) D_{\Lambda_n-1}(\delta+b) \right) =0. 
\ee 
Without external forcing ($b=0$), this expression reduces of course to the case 
discussed in the previous section; see (\ref{dc}) and (\ref{dd}). On the other hand, 
without dry friction ($\delta=0$), we end up with the integer 
spectrum of the Ornstein-Uhlenbeck process, since the Wronskian for 
the fundamental system $(D_\nu(z),D_\nu(-z))$ 
results in the identity \cite{MaObSo:66,Buch:69}
\be
D_\nu(z) D_{\nu-1}(-z)+ D_\nu(-z) D_{\nu-1}(z)  
=\sqrt{\pi}/\Gamma(-\nu+1).
\ee 
Last but not least, we stress that the eigenvalues are even functions of the driving
force $b$, since (\ref{fb}) is invariant under a change of sign. As a result, it is
sufficient to consider the case of nonnegative driving $b$.

\begin{figure}[t]
\centerline{\includegraphics[width=0.9\linewidth]{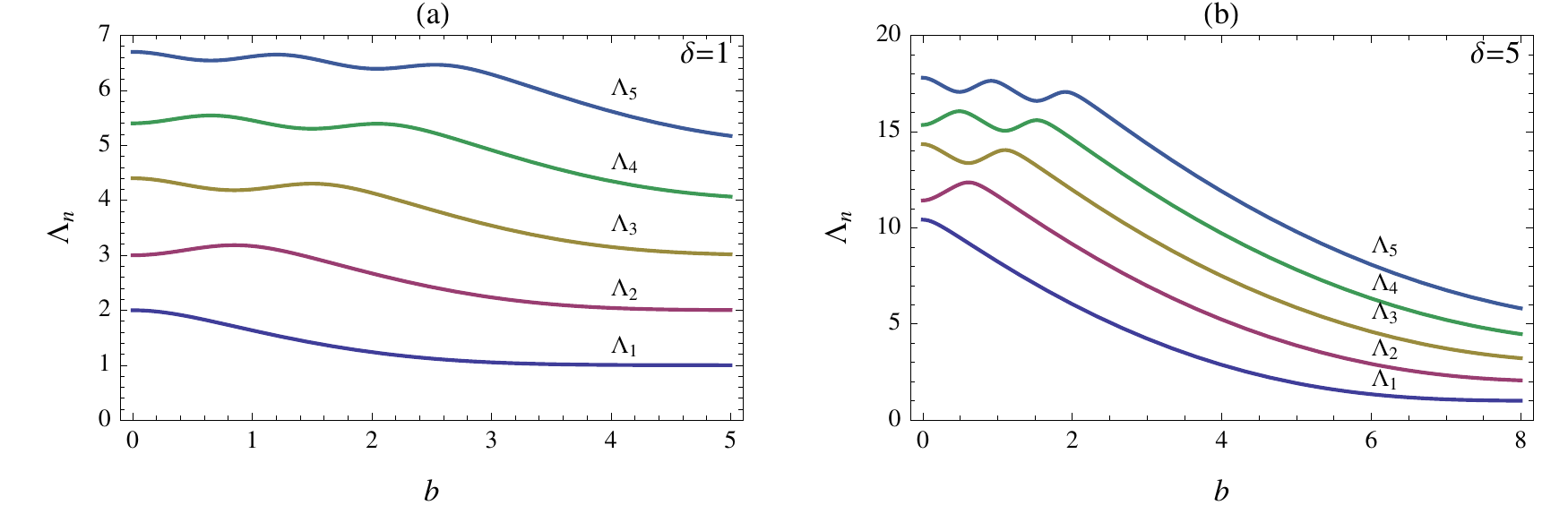}}
\caption{(Color online) Dry friction with viscous friction and forcing. 
First five eigenvalues $\Lambda_n$ as a function of $b$ for $\delta=1$ (a) and $\delta=5$ (b).}
\label{figeigenforced1}
\end{figure}

Figure \ref{figeigenforced1} illustrates the main properties of the spectrum
for fixed positive values of $\delta$. For large
forcing, i.e., in the slip phase, the spectrum approaches the integer eigenvalues 
of the Ornstein-Uhlenbeck process,
reflecting the parabolic minimum of the potential (see figure
\ref{figinvdistforced1}).
For small forcing and relatively small viscous damping, we observe the spectral 
gap between the ground state and the first nonzero eigenvalue, encountered
before. In the transition regime between stick and slip phase, 
the eigenvalues show clear oscillations
and a sharp drop when the transition value $|b|=\delta$ is approached. The drop becomes more
pronounced in the low noise limit, i.e., for large values of $\delta$ and $b$, and can be
understood from (\ref{fb}) using our knowledge of
parabolic cylinder functions gained in the previous section. 

To be more precise, consider the stick phase, i.e., $0\leq b<\delta$. 
As we have seen previously, the parabolic cylinder function
$D_\Lambda(z)$ is positive if the index $\Lambda$ is smaller 
than the lowest non-trivial eigenvalue of the corresponding undriven system, i.e., 
$\Lambda\lesssim z^2/4$ for large positive argument $z$
(see figure \ref{fig1}). Thus all contributions to (\ref{fb}) are positive if 
$\Lambda \lesssim (\delta-b)^2/4$, and this condition 
yields a lower bound for the lowest non-trivial eigenvalue, which determines, in turn,
the parabolic shape of the spectrum in the stick phase
visible in figure \ref{figeigenforced1}. This reasoning
can be turned into an asymptotic expansion for the lowest eigenvalues
if one recalls that a change in sign of one of the factors in (\ref{fb})
almost inevitably results in a zero of (\ref{fb}), given that the amplitudes 
of the terms become exponentially large.

In the slip phase $b>\delta$, the potential (\ref{fa}) has a quadratic minimum
(see figure \ref{figinvdistforced1}), and one expects the lower part of the
spectrum to be of Ornstein-Uhlenbeck type. Such a feature can be derived
from the characteristic equation (\ref{fb}). If the argument $\delta-b$ in 
(\ref{fb}) is negative, then the cylinder functions change sign at values for 
$\Lambda$ close to even nonnegative integers (see, e.g.,
(\ref{ddc}) for the lowest value or figure \ref{fig1a}). 
Since $D_\Lambda(\delta+b)$ considerably exceeds $D_{\Lambda-1}(\delta+b)$
in magnitude, the first non-trivial root of (\ref{fb}) must then occur close to one, as
observed in figure \ref{figeigenforced1}. By elaborating further
on this argument, one can
derive asymptotic expressions for the spectrum in the case $\delta>b$.

We now turn to the eigenfunctions, which for nonvanishing eigenvalues are given by the expression 
\be\label{fc} 
u_n(x)=\exp(-\Phi(x)/2) \times 
\left\{ 
\begin{array}{lll} 
D_{\Lambda_n}(x+\delta-b), & & x\geq 0 \\ 
\chi_n D_{\Lambda_n}(-x+\delta+b), & & x\leq 0 
\end{array} \right. 
\ee 
with the amplitude ratio 
\be\label{fd} 
\chi_n = \frac{D_{\Lambda_n}(\delta-b)}{D_{\Lambda_n}(\delta+b)} 
= - \frac{D_{\Lambda_n-1}(\delta-b)}{D_{\Lambda_n-1}(\delta+b)}. 
\ee 
The adjoint eigenfunctions $v_n(x)$ are as usual obtained from (\ref{ad}). 
The normalisation (\ref{af}) can conveniently be expressed as 
\be\label{fe} 
Z_n =  \Lambda_n D_{\Lambda_n}(\delta-b) D_{\Lambda_n-1}(\delta-b) \partial_{\Lambda_n} \ln \left|
\frac{D_{\Lambda_n-1}(\delta-b) D_{\Lambda_n}(\delta+b)}{D_{\Lambda_n}(\delta-b)
D_{\Lambda_n-1}(\delta+b)} \right| .
\ee 
Thus, (\ref{fc}), (\ref{fd}), and (\ref{fe}) provide the required input for the propagator 
\be\label{ff} 
p(x,\tau|x',0) = \rho_*(x) + \sum_{n=1}^\infty \exp(-\Lambda_n \tau)  
\frac{u_n(x) v_n(x')}{Z_n}. 
\ee 

\begin{figure}[t]
\centerline{\includegraphics[width=0.45\linewidth]{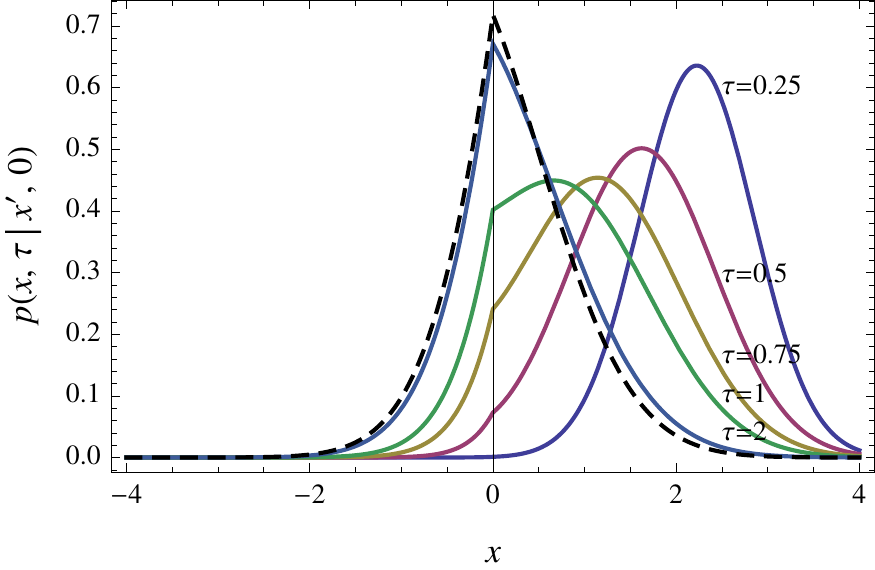}}
\caption{(Color online) Dry friction with viscous friction and forcing. Evolution of 
$p(x,\tau|x',0)$ towards the stationary distribution (dashed line) as $\tau$ increases 
for the case where $b<\delta$. Parameters: $x'=3$, $b=0.5$, and $\delta=1$.}
\label{figpropforced1}
\end{figure}

Figure \ref{figpropforced1} illustrates the time evolution of (\ref{ff}) in the stick phase
$|b|<\delta$. Qualitatively, the dynamics is quite similar to the undriven system
(see figure \ref{figpropvisc1}), with the difference that the symmetry 
of the stationary state has disappeared.

\begin{figure}[t]
\centerline{\includegraphics[width=0.9\linewidth]{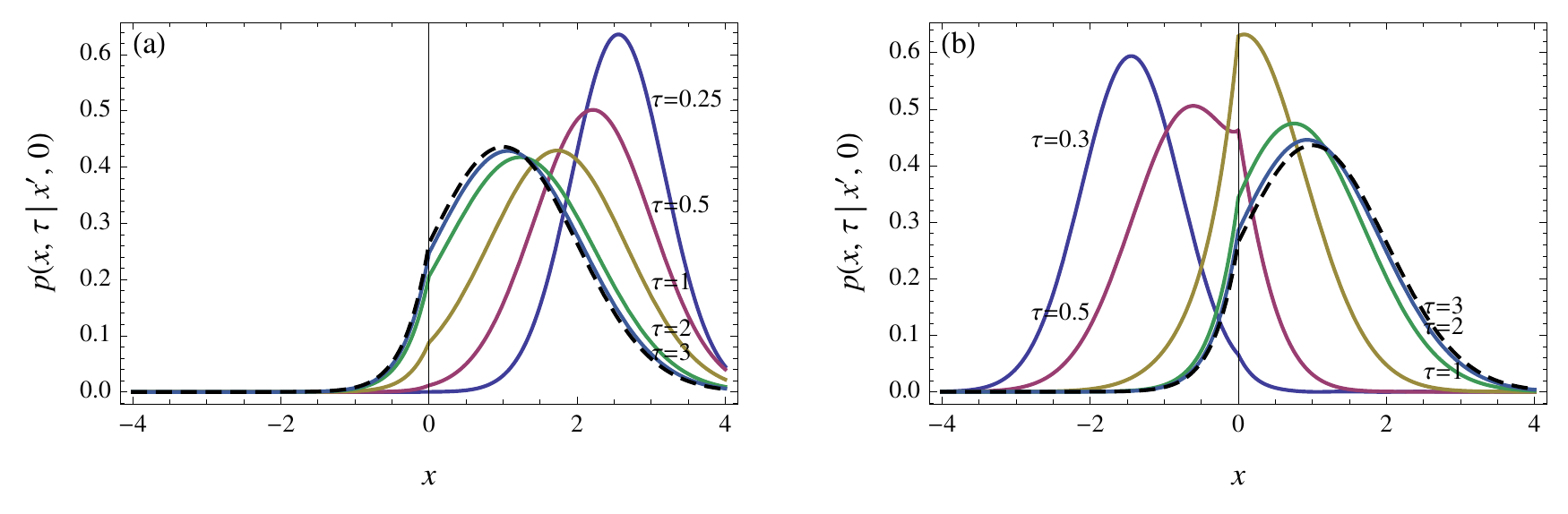}}
\caption{(Color online) Dry friction with viscous friction and forcing. Evolution 
of $p(x,\tau|x',0)$ towards the stationary distribution (dashed line) as $\tau$ increases 
for the case where $b>\delta$ starting 
from $x'=3$ (a) or $x'=-3$ (b). 
Parameters: $b=2$ and $\delta=1$.}
\label{figpropforced2a}
\end{figure}

New features appear in the slip phase $b>\delta>0$. 
Figure \ref{figpropforced2a} shows the time evolution of the propagator with
two initial conditions of opposite sign. Depending on the sign of the initial condition 
relative to the stationary state, the propagator
either remains unimodal or temporarily develops a bimodal shape. The latter is caused
by a slow process near the origin, which was already responsible for the slow build-up of
the tails of the stationary distribution in the undriven case.

\begin{figure}[t]
\centerline{\includegraphics[width=0.9\linewidth]{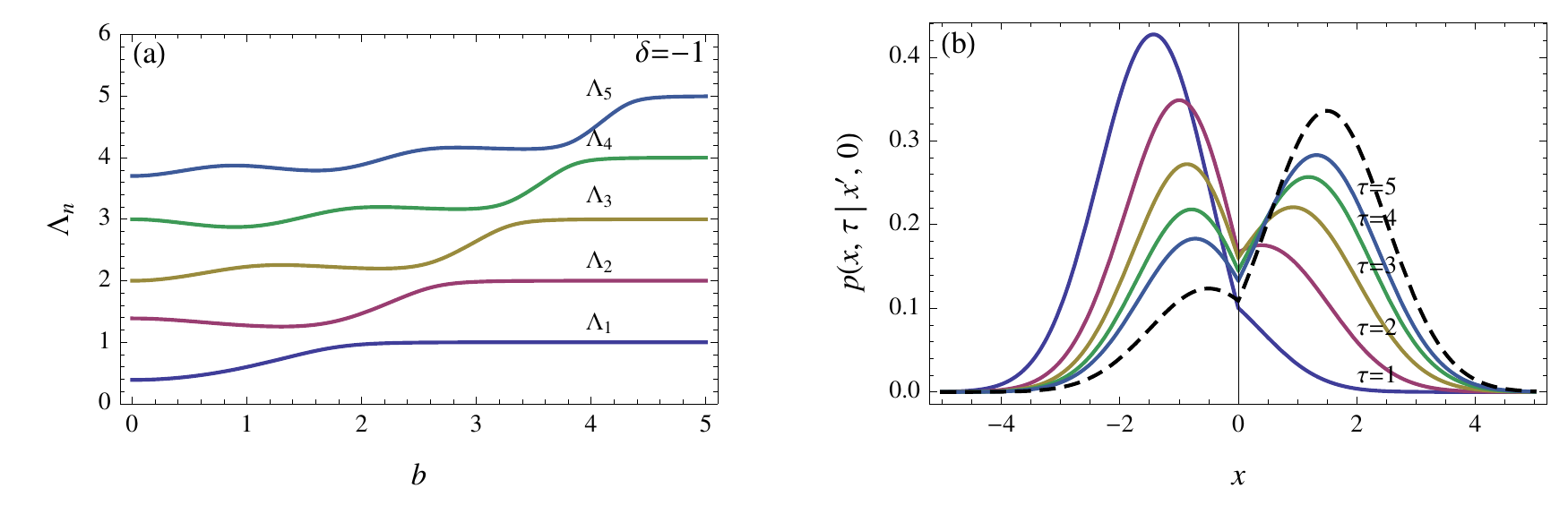}}
\caption{(Color online) Dry friction with viscous friction and  $\delta<0$. (a) First five eigenvalues as a function 
of $b$ for $\delta=-1$. (b) Evolution of $p(x,\tau|x',0)$ towards the stationary 
distribution (dashed line) as $\tau$ increases. Parameters: $x'=-3$, $\delta=-1$, and $b=0.5$.}
\label{figpropforced2}
\end{figure}

The bimodal form of $p(x,\tau|x',0)$ is also seen for negative values of $\delta$, associated
with Kramer's escape problem. Figure \ref{figpropforced2}
gives an idea of the spectrum and the time evolution of the propagator in that case.
One observes again a transition phenomenon in the spectrum when the shape of the potential 
changes from a bimodal to a unimodal structure. For $\delta<0$, the driving force $b$
affects the escape rate in the usual way. The lowest eigenvalue increases with 
increasing driving force as the potential barrier decreases in size.
Moreover, the propagator displays, as expected, a slow tunnelling process in a double-well potential.

A few more insights into the forced case can be obtained by computing the correlation function
along the lines of the previous section; see, e.g., (\ref{di}) and (\ref{dj}). In the present case, 
(\ref{fc})  
yields, using (\ref{efa}) and integration by parts, 
\begin{eqnarray}
\int_{-\infty}^\infty x\, u_n(x)\, dx &=& e^{-\delta^2/4}[D_{\Lambda_n-2}(\delta-b)-
\chi_\Lambda D_{\Lambda_n-2}(\delta+b)]\nonumber\\
&=& \frac{2 \delta  e^{-\delta^2/4}}{\Lambda_n-1} D_{\Lambda_n-1}(\delta-b),
\label{fg} 
\end{eqnarray} 
where for the last step the recurrence relation (\ref{djb}) and the definition (\ref{fd})
have been used. From the propagator (\ref{ff}), we then obtain
\begin{eqnarray}
\label{fh} 
\langle x(\tau) x(0)\rangle - \langle x\rangle_*^2 &=&
\sqrt{\frac2\pi} \frac{4 \delta^2}
{e^{(\delta-b)^2/2}\,\erfc\left(\frac{\delta-b}{\sqrt{2}}\right)+ 
e^{(\delta+b)^2/2}\,\erfc\left(\frac{\delta+b}{\sqrt{2}}\right)} \nonumber \\
& & \times 
\sum_{n=1}^\infty \frac{\exp(-\Lambda_n \tau)}{\Lambda_n (\Lambda_n-1)^2} 
\frac{D_{\Lambda_n-1}(\delta-b)}{D_{\Lambda_n}(\delta-b) 
\partial_{\Lambda_n} \ln \left|
\frac{D_{\Lambda_n-1}(\delta-b) D_{\Lambda_n}(\delta+b)}{D_{\Lambda_n}(\delta-b)
D_{\Lambda_n-1}(\delta+b)} \right| }\, .
\end{eqnarray}
In the limit $b\rightarrow 0$, this result reduces
to the correlation function of the undriven case, (\ref{dja}), as the
denominator of the expansion coefficient tends towards
$-2\partial_{\Lambda_n}D_{\Lambda_n}(\delta)$ and the numerator ensures that
only the odd part of the spectrum contributes; see (\ref{dd}).

\begin{figure}[t]
\centerline{\includegraphics[width=0.45\linewidth]{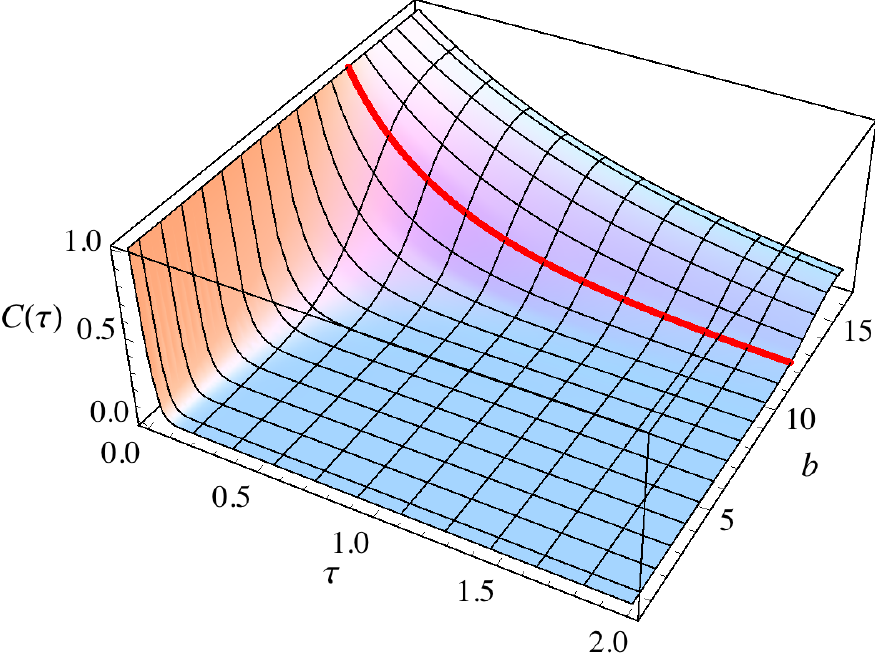}}
\caption{(Color online) Dry friction with viscous friction and forcing.
Normalised correlation function 
$C(\tau)=(\langle x(\tau) x(0)\rangle - \langle x\rangle_*^2)/(\langle x^2\rangle_*-\langle x \rangle_*^2)$
as a function of $\tau$ and $b$ for $\delta=12$. The red line at $b=\delta$ marks the transition
between the stick and slip states.}
\label{corr2}
\end{figure}

Figure \ref{corr2} shows the plot of the correlation function normalised by the stationary variance
as a function of $\tau$ and $b$. The red line in this plot marks the boundary $b=\delta$ separating
the noiseless or deterministic stick and slip states. In the presence of noise, these two states
are now recognisable as two different correlation phases. In the stick phase, $0<b<\delta$,
it can be shown from the spectrum of the Fokker-Planck operator
that the correlation time governing the exponential decay of $\langle x(\tau) x(0)\rangle$ scales 
according to $(\delta-b)^{-2}$, and so dramatically increases near the stick-slip transition.
In the slip phase ($b> \delta$), by contrast, the correlation time is mostly determined by the
viscous damping, and thus does not change sensibly as a function of $b$. This difference in the behaviour of 
$\langle x(\tau) x(0)\rangle$ is clearly seen in figure \ref{corr2} and becomes more visible in the small-noise limit.

\section{Conclusions} 
\label{seccon} 
 
In this paper we have studied a Langevin equation that includes a solid or dry friction force and an external
constant force in addition to the viscous friction force commonly considered in the context of Brownian motion. 
This stochastic equation extends an earlier model studied by de~Gennes \cite{gennes2005}, which
can be thought of as describing the dynamics of a solid object resting on a tilted 
solid surface, which is vibrated randomly with Gaussian white noise. By solving the time-dependent
Fokker-Planck equation associated with this model, we have obtained the time-dependent
propagator, which gives the probability that the object has a certain velocity at any time starting from
a given initial velocity, in addition to the velocity correlation function.

These results, combined with the full spectrum of the Fokker-Planck operator, allowed us to 
obtain a detailed understanding of dry friction in the presence of noise. 
In particular, we have seen that the singular nature of the dry friction force at zero velocity gives rise to
a cusp in the propagator whenever the external forcing is smaller in magnitude than the dry friction
force. We have also seen that the viscous damping does not alter the properties of the model
much when the noise is sufficiently small, essentially because the stochastic dynamics is then
confined near the origin (zero velocity state) which is most sensitive to dry friction. Conversely,
for large noise powers, the model is mostly dominated by diffusion and behaves
similarly as an Ornstein-Uhlenbeck process. Finally, we have
seen that, although the stick-slip transition is blurred at the level of the stochastic dynamics, it is possible
to define stick and slip phases at the level of the correlation function. The stick phase is characterized by a sharp
increase of the correlation time with the external force, whereas, for the slip phase, the correlation time
is mostly determined by diffusion and is almost independent of the external force. These
quantitative predictions could be checked in real experiments and in more elaborate models 
of dry friction. 

To conclude, it is worth noting that the results presented here have direct analogs in terms of 
quantum mechanical potentials. Indeed, some of our results about the spectrum of the 
Fokker-Planck operator in the presence of dry and viscous frictions can be inferred from 
previously-published results on the so-called 
constrained quantum harmonic oscillator \cite{Dean_PCPS66,auluck1945}. However, to the 
best of our knowledge, the solution that we have obtained for the full-force case has not
been considered before. It might be of interest to translate this solution to the quantum
context. Furthermore, it might be interesting, at least from an aesthetic point of view, to see
whether the spectral sums obtained for dry and viscous forces and the full-force case can be written in 
closed analytical forms, as in the case of pure dry friction. This might be possible given that the 
Fokker-Planck equation is piecewise linear in all cases. 
 
\begin{acknowledgments}
This work was supported by an RCUK Interdisciplinary Fellowship (H.T.),
and EPSRC grants nos.\ EP/E049257/1 (E.V.d.S.) and EP/H04812X/1 (W.J.).
\end{acknowledgments}

\appendix 
  
\section{Particle in a wedge} 
\label{secwed} 
 The stochastic dynamics of a particle in a symmetric wedge potential 
\be\label{ca} 
\Phi(x)=|x| 
\ee 
is a classical textbook example of a Fokker-Planck operator with spectral gap 
and continuous spectrum \cite{Risk:89}. Here we recall the solution of this well-known and
somewhat elementary problem, as it will certainly help the interested reader to understand the
more elaborate considerations found in the other appendices. 

Eigenfunctions are either even, 
$u^{(e)}_\Lambda(-x)= u^{(e)}_\Lambda(x)$, or  
odd, $u^{(o)}_\Lambda(-x)= -u^{(o)}_\Lambda(x)$, with respect to inversion. 
The eigenvalue equation (\ref{ac}) may be considered on the positive domain only  
and the boundary condition, given before in (\ref{agb}), readily reduces to 
\be\label{cb} 
u^{(e)}_\Lambda(0) + \left.\frac{d u^{(e)}_\Lambda (x)}{dx}\right|_{x=0} =0, \qquad 
u^{(o)}_\Lambda(0)=0 \, . 
\ee 
The general solution of (\ref{ac}) can be expressed  
as a linear combination of exponentials.  
Apart from the stationary distribution (\ref{ab}) with eigenvalue 
$\Lambda=0$, the solution which satisfies the symmetry, the boundary condition (\ref{cb}),  
and which decays to zero at infinity is given by 
\begin{eqnarray}
u^{(e)}_\Lambda(x) &=&e^{-x/2}\,[2 \beta_\Lambda \cos(\beta_\Lambda x) - \sin(\beta_\Lambda x)] \nonumber\\
u^{(o)}_\Lambda(x) &=& e^{-x/2} \sin(\beta_\Lambda x)\label{cc} 
\end{eqnarray}
for $x\geq 0$, where 
\be\label{cd} 
\beta_\Lambda=\sqrt{\Lambda-1/4} >0. 
\ee 
The latter inequality determines the spectral gap and 
restricts the continuous spectrum to the range $\Lambda>1/4$. 
As for the normalisation of the odd eigenmodes, (\ref{cc}) and (\ref{ad}) result in 
\begin{eqnarray}\label{ce} 
\int_{-\infty}^\infty v^{(o)}_{\Lambda'}(x)\, u^{(o)}_\Lambda(x)\, dx &=& \int_0^\infty \left[  
\cos((\beta_\Lambda-\beta_{\Lambda'})x) - \cos((\beta_\Lambda+\beta_{\Lambda'})x) \right]\, dx 
\nonumber \\ 
&=& 2 \pi \sqrt{\Lambda-1/4}\, \delta(\Lambda-\Lambda') = Z^{(o)}_\Lambda  \delta(\Lambda-\Lambda'),
\end{eqnarray} 
whereas for the even eigenmodes, we obtain 
\begin{eqnarray}
\label{cf} 
\int_{-\infty}^\infty v^{(e)}_{\Lambda'}(x)\, u^{(e)}_\Lambda(x)\, dx &=& 
\int_0^\infty \left[ 
(4 \beta_\Lambda \beta_{\Lambda'} +1) 
\cos((\beta_\Lambda -\beta_{\Lambda'})x) \right. \nonumber\\  
& & \quad+ (4 \beta_\Lambda \beta_{\Lambda'} -1) 
\cos((\beta_\Lambda +\beta_{\Lambda'})x) 
+2 (\beta_\Lambda - \beta_{\Lambda'}) 
\sin((\beta_\Lambda - \beta_{\Lambda'}) x) \nonumber \\
& & \quad\left. -2 (\beta_\Lambda + \beta_{\Lambda'}) 
\sin((\beta_\Lambda + \beta_{\Lambda'}) x) \right]\, dx \nonumber \\ 
&=& 2 \pi (4 \beta_\Lambda^2 +1) \sqrt{\Lambda-1/4}\, 
\delta(\Lambda-\Lambda')\nonumber\\ 
&=& Z^{(e)}_\Lambda  \delta(\Lambda-\Lambda'). 
\end{eqnarray} 
The contribution of the odd eigenmodes to the propagator (\ref{ae}) can 
now be evaluated 
straightforwardly using (\ref{cc}), (\ref{ad}) and (\ref{ce}):
\begin{eqnarray}
\label{cg} 
p^{(o)}(x,\tau|x',0) &=& \int_{1/4}^\infty e^{-\Lambda \tau}\,  
u^{(o)}_\Lambda(x)\, v^{(o)}_\Lambda(x') /Z^{(o)}_\Lambda\, 
d \Lambda \nonumber \\ 
&=& \frac{e^{-\tau/4} 
e^{-(|x|-|x'|)/2}}{2 \pi} \int_0^\infty e^{-\beta^2 \tau} \left[
\cos(\beta (x-x')) - \cos(\beta (x+x')) \right] d \beta \nonumber \\ 
&=&  \frac{e^{-\tau /4} e^{-(|x|-|x'|)/2}}{4 \sqrt{\pi \tau}} 
\left( e^{-(x-x')^2/(4 \tau)} 
- e^{-(x+x')^2/(4 \tau)} \right).
\end{eqnarray} 
For the even modes, we obtain instead
\begin{eqnarray}\label{ch} 
p^{(e)}(x,\tau|x',0) &=& \int_{1/4}^\infty e^{-\Lambda \tau}\, 
u^{(e)}_\Lambda(x)\,  v^{(e)}_\Lambda(x') /Z^{(e)}_\Lambda\, 
d \Lambda\nonumber \\ 
&=& \frac{e^{-\tau/4} e^{-(|x|-|x'|)/2}}{2 \pi} \int_0^\infty 
\frac{e^{-\beta^2 \tau}}{4 \beta^2+1} \left[ (4 \beta^2+1)
\cos(\beta(|x|-|x'|)) \right.\nonumber \\ 
& & \quad \left.+ 
(4 \beta^2-1)\cos(\beta(|x|+|x'|)) -4 \beta 
\sin(\beta(|x|+|x'|)) \right] d \beta. 
\end{eqnarray} 
The remaining non-trivial integrals can be evaluated using the 
identities \cite{GrRy:80} 
\begin{eqnarray}
\label{ci} 
\int_0^\infty e^{-\beta x^2}\sin(ax) \frac{x\, dx}{\gamma^2 + x^2} 
&=& -\frac{\pi}{4} e^{\beta \gamma^2}\left[ 2 \sinh(a\gamma) 
+ e^{-\gamma a} \erf\left(\frac{2 \gamma\beta-a}{2\sqrt{\beta}}\right) 
- e^{\gamma a} \erf\left(\frac{2 \gamma\beta+a}{2\sqrt{\beta}}\right) 
\right]  \label{cia}  \\ 
\int_0^\infty e^{-\beta x^2}\cos(ax) \frac{dx}{\gamma^2 + x^2} 
&=& \frac{\pi}{4\gamma} e^{\beta \gamma^2}\left[ 2 \cosh(a\gamma) 
- e^{-\gamma a} \erf\left(\frac{2 \gamma\beta-a}{2\sqrt{\beta}}\right) 
- e^{\gamma a} \erf\left(\frac{2 \gamma\beta+a}{2\sqrt{\beta}}\right) 
\right]. 
\label{cib} 
\end{eqnarray} 
We thus end up with\footnote{See \cite{Wong_AMS64} for a closely related result.}
\be
\label{cj} 
p^{(e)}(x,\tau|x',0) = 
\frac{e^{-\tau/4} e^{-(|x|-|x'|)/2}}{4 \sqrt{\pi \tau}} 
\left( e^{-(x-x')^2/(4\tau)} +  e^{-(x+x')^2/(4\tau)} \right)  
-\frac{e^{-|x|}}{4} \left[1- 
\erf\left(\frac{\tau-(|x|+|x'|)}{2\sqrt{t}}\right) \right].
\ee
The final result, (\ref{bb}), then follows from (\ref{cg}), 
(\ref{cj}) and 
\be\label{ck} 
p(x,\tau|x',0)=\rho_*(x)+p^{(e)}(x,\tau|x',0)+p^{(o)}(x,\tau|x',0) . 
\ee 
 
\section{Symmetric piecewise parabolic potential} 
\label{secsym} 
 
For the symmetric potential (\ref{db}), it is sufficient to consider the eigenvalue equation 
(\ref{ac}) for nonnegative argument with appropriate boundary condition at the origin, since 
eigenfunctions are either even or odd. The boundary condition expressed in 
(\ref{agb}) here yields 
(see (\ref{cb}))
\be\label{ea} 
\delta u^{(e)}_\Lambda(0) + \left.\frac{d u^{(e)}_\Lambda (x)}{dx}\right|_{x=0} =0, \qquad 
u^{(o)}_\Lambda(0)=0 \, . 
\ee 
The differential equation (\ref{ac}) with the potential (\ref{db}) is a 
special case of the so-called
Kummer's equation, whose solution can be expressed in terms of 
Kummer functions. 
The solution which decays at infinity can conveniently be expressed 
in terms the parabolic cylinder 
functions.\footnote{We use Whittaker's notation; see, e.g., 
\cite{AbSt:70}}
To be more precise, we have that
\be
u(z)=e^{-z^2/4}\, D_\nu(z),
\ee 
solves 
\be
u''+(zu)'+\nu u=0,
\ee 
and $D_\nu(z) \sim e^{-z^2/4}\, z^{\nu}$ as $z\rightarrow +\infty$. 
Thus the solution of (\ref{ac}), which decays at infinity, is given by 
\be\label{eb} 
u_\Lambda(x)=\exp(-(x+\delta)^2/4)\, D_\Lambda(x+\delta), \qquad x\geq 0.
\ee 

The spectrum is now determined by the boundary condition at the origin. 
For the odd eigenfunctions, (\ref{ea}) and (\ref{eb}) yield the condition (\ref{dc}). 
If we use the following differential identity for the parabolic cylinder function: 
\be\label{ec} 
D'_\Lambda(z)+\frac{z}{2} D_\Lambda(z)=\Lambda D_{\Lambda-1} (z), 
\ee 
then the boundary condition for the even eigenfunctions results in 
\be\label{ed} 
\Lambda_n^{(e)} D_{\Lambda_n^{(e)}-1} (\delta)=0. 
\ee 
This means that, apart from the trivial zero eigenvalue associated with the stationary state, all the 
other eigenvalues obey (\ref{dd}) if we take the result of (\ref{dc}) into account. 
 
Turning now to the eigenfunctions, it should be clear that (\ref{eb}) leads to (\ref{de}). 
The adjoint eigenfunctions are determined by (\ref{ad}). The orthogonality 
of the eigenfunctions follows from the standard differential identities for the 
parabolic cylinder function, which are expressed in (\ref{ec}) and can be written as 
\begin{eqnarray}
\frac{d \exp(-z^2/4) D_\nu(z)}{d z} &=& -\exp(-z^2/4) D_{\nu+1}(z) 
\label{efa} \\
\frac{d \exp(z^2/4) D_\nu(z)}{d z} &=& \nu \exp(z^2/4) D_{\nu-1}(z) 
\label{efb}. 
\end{eqnarray} 
As for the normalisation, using (\ref{efa}) and (\ref{efb}), we obtain the relations 
\be\label{eg} 
\frac{ d u_n^{(o)}(x)}{dx} = -u_n^{(e)}(x), \qquad 
\frac{ d v_n^{(e)}(x)}{dx} = \Lambda_n^{(e)} v_n^{(o)}(x),
\ee 
relating the eigenfunctions and the eigenfunctions of the adjoint problem. 
Integration by parts thus leads to 
\be\label{eh} 
\int_0^\infty v_n^{(e)}(x)\, u_n^{(e)}(x)\, dx = \Lambda_n^{(e)} 
\int_0^\infty v_n^{(o)}(x)\, u_n^{(o)}(x)\, dx. 
\ee 

This last equation yields (\ref{dg}). In fact, (\ref{eg}) 
are the remnants of the differential relations for Hermite polynomials,
which guarantee that for the case $\delta=0$,  i.e., for the Ornstein-Uhlenbeck 
process, all eigenfunctions can be obtained by the
shift operation.

To evaluate the normalisation integrals and to establish the orthogonality of
eigenfunctions, we resort to a known integral identity of parabolic
cylinder functions which follows from (\ref{efa}) and (\ref{efb}) 
using integration by parts \cite{Buch:69}:
\begin{eqnarray}\label{ei}
(\Lambda-\Lambda') \int_a^\infty D_\Lambda(x) 
D_{\Lambda'}(x) dx
&=& \int_a^\infty \left(\frac{d \exp(x^2/4) D_{\Lambda+1}(x)}{dx}\exp(-x^2/4) 
D_{\Lambda'}(x)\right. \nonumber \\
& & \quad\left. - \exp(x^2/4) D_\Lambda(x) \frac{d \exp(x^2/4) D_{\Lambda'+1}(x)}{dx}\right) dx  
\nonumber \\
&=& D_\Lambda(a) D_{\Lambda'+1}(a) - D_{\Lambda+1}(a) D_{\Lambda'}(a) .
\end{eqnarray}
Choosing $a=\delta$ and $\Lambda^{(o)}_n=\Lambda \neq \Lambda'=\Lambda^{(o)}_m$, the
right hand side of (\ref{ei}) vanishes due to the characteristic
equation (\ref{dc}) and the left hand side results in the orthogonality
condition
\be\label{ee} 
\int_{-\infty}^\infty v_m^{(o)}(x)\, u_n^{(o)}(x)\, dx = 2 \int_0^\infty  
D_{\Lambda_m^{(o)}}(x+\delta)  D_{\Lambda_n^{(o)}}(x+\delta)\, dx=0, \qquad m\neq n 
\ee 
for parabolic cylinder functions with index determined by (\ref{dc}). 
This relation is expected, as eigenfunctions are supposed to be mutually orthogonal. 

To evaluate the normalisation integral, let us consider the case $a=\delta$ and 
$\Lambda=\Lambda_n^{(o)}$. Then (\ref{ei}) results in
\be\label{ej}
\int_\delta^\infty D_{\Lambda_n^{(o)}}(x) D_{\Lambda'}(x) dx
=  D_{\Lambda_n^{(o)}+1}(\delta)
\frac{D_{\Lambda_n^{(o)}}(\delta)-D_{\Lambda'}(\delta)}{\Lambda_n^{(o)}-\Lambda'}
\ee
if we again take the characteristic equation into account. In the limit
$\Lambda'\rightarrow \Lambda^{(o)}_n$, we then obtain the result 
(\ref{df}) using the recurrence relation
\be\label{djb}
zD_\nu(z)-D_{\nu+1}(z)-\nu D_{\nu-1}(z)=0 \, .
\ee 

\section{Biased forcing} 
\label{secbia} 
 
For the potential (\ref{fa}), the eigenvalue equation (\ref{ac}) reads 
\be\label{ga} 
u''_\Lambda(x) + ((x \pm \delta-b)u_\Lambda)'+\Lambda u_\Lambda(x) =0,
\ee 
where the $+$ sign applies for $x>0$ and the $-$ sign applies for $x<0$. 
The solution that fulfills the boundary condition at infinity can be written again
in terms of parabolic cylinder functions: 
\be\label{gb} 
u_\Lambda(x)=C_\pm \exp(-(x\pm \delta-b)^2/4) D_\Lambda(\pm x+\delta \mp b), \qquad x \gtrless 0. 
\ee 
The boundary conditions (\ref{agb}) thus result in 
\begin{eqnarray}
\label{gc} 
C_- \exp(-(\delta+b)^2/4) D_{\Lambda}(\delta+b) &=&
C_+ \exp(-(\delta-b)^2/4) D_{\Lambda}(\delta-b) 
\label{gca}\\  
C_- \Lambda \exp(-(\delta+b)^2/4) D_{\Lambda-1}(\delta+b) &=&
-C_+ \Lambda \exp(-(\delta-b)^2/4) D_{\Lambda-1}(\delta-b). 
\label{gcb} 
\end{eqnarray} 
The condition for a non-trivial solution yields the characteristic equation (\ref{fb}). 
As for the amplitudes $C_\pm$, there does not seem to be a convenient form which can 
cope both with the limit of symmetric potentials and the different symmetries of the eigenfunctions. 
If we choose $C_+=\exp(b^2/4-\delta b/2)$, then (\ref{gb}), (\ref{gca}), and (\ref{gcb}) lead to 
(\ref{fc}) and (\ref{fd}).  
 
The normalisation (\ref{af}) of the eigenfunctions (\ref{fc}) and (\ref{ad}) 
reads
\be
Z_n= \int_0^\infty D_{\Lambda_n}(x+\delta-b)  D_{\Lambda_n}(x+\delta-b) dx
+ \chi_n^2
\int_0^\infty D_{\Lambda_n}(x+\delta+b)  D_{\Lambda_n}(x+\delta+b) dx.
\label{gd}
\ee
Using the identity (\ref{ei}) with $a=\delta \pm b$, $\Lambda=\Lambda_n$ and
$\Lambda' \rightarrow \Lambda_n$, we can express the remaining integrals 
in terms of parabolic cylinder functions to obtain
\begin{eqnarray}
\label{ge}
Z_n &=& D_{\Lambda_n+1}(\delta- b) \partial_{\Lambda_n} 
D_{\Lambda_n}(\delta- b)
- D_{\Lambda_n}(\delta- b) \partial_{\Lambda_n} D_{\Lambda_n+1}(\delta- b) \nonumber\\
& & \quad + \chi_n^2
\left[D_{\Lambda_n+1}(\delta+ b) \partial_{\Lambda_n} D_{\Lambda_n}(\delta+ b)
- D_{\Lambda_n}(\delta+ b) \partial_{\Lambda_n} 
D_{\Lambda_n+1}(\delta+ b)\right].
\end{eqnarray}
Using the identity (\ref{djb}) to eliminate the cylinder functions with the largest index,
$D_{\Lambda_n+1}(\delta\pm b)$, and applying the definition (\ref{fd}),
we finally arrive at the result of (\ref{fe}).

\bibliography{touchettedryfriction1} 
 
\end{document}